\documentclass[preprint]{aastex6}

\begin{document}


\title{Characteristics that Produce White-Light Enhancements in Solar Flares Observed by \textit{Hinode}/SOT}


\author{Kyoko Watanabe\altaffilmark{1}}
\affil{National Defense Academy of Japan, \\
1-10-20 Hashirimizu, Yokosuka 239-8686, Japan}

\and

\author{Jun Kitagawa, Satoshi Masuda}
\affil{Institute for Space-Earth Environmental Research (ISEE), Nagoya University, \\
Furo-cho, Chikusa-ku, Nagoya 464-8601, Japan}


\altaffiltext{1}{kwatana@nda.ac.jp}

\begin{abstract}
To understand the conditions that produce white-light (WL) enhancements in solar flares, a statistical analysis of visible continuum data as observed by \textit{Hinode}/Solar Optical Telescope (SOT) was performed.
In this study, approximately $100$ flare events from M- and X-class flares were selected.
The time period during which the data were recorded spans from January 2011 to February 2016.
Of these events, approximately half are classified as white-light flares (WLFs), whereas the remaining events do not show any enhancements of the visible continuum (non-WLF; NWL).
In order to determine the existence of WL emission, running difference images of not only the \textit{Hinode}/SOT WL (G-band, blue, green, and red filter) data but also the \textit{Solar Dynamics Observatory}/Helioseismic and Magnetic Imager continuum data are used.
A comparison between these two groups of WL data in terms of duration, temperature, emission measure of \textit{GOES} soft X-rays, distance between EUV flare ribbons, strength of hard X-rays, and photospheric magnetic field strength was undertaken.
In this statistical study, WLF events are characterized by a shorter time-scale and shorter ribbon distance compared with NWL events.
From the scatter plots of the duration of soft X-rays and the energy of non-thermal electrons, a clear distinction between WLF and NWL events can be made.
It is found that the precipitation of large amounts of accelerated electrons within a short time period plays a key role in generating WL enhancements.
Finally, it was demonstrated that the coronal magnetic field strength in the flare region is one of the most important factors that allow the individual identification of WLF events from NWL events.
\end{abstract}


\section{Introduction}
\label{intro}

Solar flares are often associated with enhancements of visible continuum (white-light; WL) radiation.
The first white-light flare (WLF) recorded was the Carrington flare of 1859 \citep{Carrington1859}.
WLFs are largely associated with energetic events such as \textit{GOES} X-class flares and are rarely observed.
However, using recent high-precision observations obtained from spacecraft (\textit{Yohkoh}, \textit{TRACE}, \textit{Hinode}), WLFs have been observed in weaker flares such as \textit{GOES} C-class flares \citep{Matthews2003, Hudson2006, Jess2008, Wang2009}.

Although $150$ years has passed since the discovery of WLF, the mechanism of WL emission is still not fully understood.
One of the most famous correlations with WL emission is that of hard X-ray emission, which originates from accelerated electrons.
Observationally, WL emission is well correlated with hard X-ray and radio emission, both in the time profile and emission location \citep[e.g.,][]{Neidig1989, Ding2003, Fletcher2007, Watanabe2010, Krucker2011, Kuhar2016}.
As a result there is some consensus that the origin of WL emission is non-thermal electrons.
By comparing the total energy of WL and hard X-ray emission, the energy range characterizing WL emission can be estimated as a few tens of keVs \citep{Neidig1989, Fletcher2007, Watanabe2010, Kuhar2016}.
The total energy of the observed WL emission is therefore similar to the total energy of the accelerated electrons with energies typical of hard X-rays.

There are questions relating to the emission height of WLF.
Theoretically, WL is emitted near the photosphere.
However, non-thermal electrons in the energy range of $50-100~{\rm keV}$ are almost thermalized by the time they reach the lower chromospheres, whereas hard X-rays are emitted from the lower chromosphere.
To reach the photosphere, accelerated electrons need energies in excess of $900~{\rm keV}$  \citep{Neidig1989}.
Even if such high-energy electrons exist, this is still not enough to explain the total energy of WL emission.

Observationally, the emission height of WL and hard X-rays and the relationship between them are measured by limb flares \citep[e.g.,][]{BattagliaKontar2011, Oliveros2012, BattagliaKontar2012, Watanabe2013, Krucker2015}.
Some events show that WL emission takes place in the photosphere \citep{Oliveros2012, Watanabe2013}, whereas other events show that it occurs in the chromosphere \citep{BattagliaKontar2012, Krucker2015}.
Even in the same flare, different results were reported.
For the 2011 February 24 flare, one paper reported there was a significant difference in source height between hard X-rays and WL \citep{BattagliaKontar2012}, however, others showed no difference between them with a different analysis method \citep{Oliveros2012}.
Determining the height of WL emission is therefore not a straightforward problem and at present its exact nature remains unsolved.

The emission height relationship between WL and hard X-rays reflects the emission mechanisms of WL emissions.
Theories explaining WL emission mechanisms fall into two general categories, namely, one involves direct heating and the other indirect heating.
A simple model for the direct heating case is that very high-energy ($\gg100~{\rm keV}$) electrons precipitate directly into the photosphere, thereby increasing the temperature of the photosphere and resulting in the emission of WL \citep{AboudarhamHenoux1986,Neidig1989}. 
A further model involving the direct heating approach is that WL emission results from an optically thin source in the mid-chromosphere that is directly heated by non-thermal electrons \citep{Kerr2014}.
An indirect heating model of WL emission may be outlined as follows: The WL emission region/layer differs from the energy deposition layer/region for non-thermal electrons.
Relatively low-energy ($< 100~{\rm keV}$) electrons precipitate into the chromosphere wherein they dissipate energy.
Energy is then transported from this heated region to the lower atmosphere.
This energy transport is termed “back-warming” and the exact transport mechanism remains the topic of debate \citep{Machado1989, Metcalf1990, Isobe2007}.
The resulting WL emission is thought to be caused by the photoionization of hydrogen atoms and recombination of associated photoelectrons.
The excited neutral hydrogen atoms lead to Balmer Paschen continuum emission \citep{Machado1986, Metcalf2003}. 

Although these emission mechanism models highlight the relationship between WL and hard X-ray emission, there are many flare events that do not have any WL enhancements even if they have hard X-ray emission.
There are many reports of WLFs that discuss the correlation between hard X-rays and emission mechanisms.
However, there are no studies that compare events without WL enhancements even if the flare itself is observed by continuum bands.
In order to understand the conditions that produce enhancements of WL in solar flares, a statistical analysis was performed on WL data observed by the Solar Optical Telescope \citep[SOT;][]{Tsuneta2008, Suematsu2008, Shimizu2008, Ichimoto2008} onboard \textit{Hinode} and the Helioseismic and Magnetic Imager \citep[HMI;][]{Scherrer2012, Schou2012} onboard \textit{Solar Dynamics Observatory} \citep[\textit{SDO};][]{Pesnell2012}.
We compared these WL data with the data of the \textit{GOES} soft X-rays, the \textit{Reuven Ramaty High Energy Solar Spectroscopic Imager} \citep[\textit{RHESSI};][]{Lin2002} hard X-rays, and the strength of the photospheric magnetic fields, as observed by \textit{SDO}/HMI.
An investigation of the relationships between the many physical parameters recorded was completed and the results presented in this work provide some constraints on the mechanism of WL emission.

\section{Event Selection}
\label{event}

The \textit{Hinode}/SOT provides high-resolution photometric and magnetic observations of various features in the photosphere and chromosphere and has the capability of observing WLFs.
The broadband filtergraph imager (BFI) on SOT contains interference filters to acquire images of the Ca {\sc ii} H ($3968.5~{\rm \AA}$, width $3~{\rm \AA}$), G-band ($4305.0~{\rm \AA}$, width $8~{\rm \AA}$), blue filter ($4504.5~{\rm \AA}$, width $4~{\rm \AA}$), green filter ($5550.5~{\rm \AA}$, width $4~{\rm \AA}$), and red filter ($6684.0~{\rm \AA}$, width $4~{\rm \AA}$).
From 2011, SOT performed a flare observation program that obtained continuum images of the G-band and the red, green, and blue filters when a solar flare was automatically detected \citep{Kano2008} by the X-Ray Telescope \citep[XRT;][]{Golub2007} and the flare position was inside the SOT’s field of view (FOV).

Not all flares are observed by SOT due to restrictions related to SOT’s FOV.
Among the \textit{Hinode} instruments, XRT has the capability to observe the full solar disk.
SOT has a much smaller maximum FOV size of $328" \times 164"$.
Therefore, only flares that occur while \textit{Hinode} is observing and are located inside the FOV may be observed by \textit{Hinode}.
It is important to determine which flares were observed by \textit{Hinode} in order to analyze the flare data.
Flare events that occurred while \textit{Hinode} was observing were listed and checked to determine whether the event was inside the \textit{Hinode} FOV.
These results are available on the web site in the \textit{Hinode} flare catalog \citep{Watanabe2012}.
In this catalog, the number of images obtained by the \textit{Hinode} instruments is shown along with the \textit{RHESSI} and Nobeyema radio heliograph information.

In this study, events were selected for the period January 2011 to February 2016.
In order to select flare events, the \textit{Hinode} flare catalog \citep{Watanabe2012} was used.
This catalog lists $11387$ flare events during the study period.
M- and X-class flares were chosen for investigation and this is because WLFs are usually associated with relatively large flares.
Of the total $11387$ events, $721$ events satisfied these selection criteria.
From this total of $721$, events were selected that were observed using \textit{Hinode}/SOT in the visible continuum bands (G-band, blue, green, and red filters) during flare observation mode.
This gave a revised total of $101$ events.
These $101$ events were classified into WLF events and NWL events using running difference images in the SOT continuum data.
The criterion for classification of WLF was the existence of WL enhancements under the Ca {\sc ii} H ribbon.
All images through the flare evolution were searched for a WL signature.
This resulted in the identification of $36$ WLF events and $65$ NWL events.
However, it is possible that even if an event is classified as NWL from SOT data, WL emission may exist outside the SOT’s FOV.
In order to account for this factor, the \textit{SDO}/HMI continuum data were checked and $13$ further WLF events were identified (these WL enhancements were located on the $1600~{\rm \AA}$ ribbons).
The final sample consisted of $49$ WLF events ($11$ X-class and $38$ M-class flares) and $52$ NWL events ($5$ X-class and $47$ M-class flares).
The event lists for WLF and NWL events are given in Tables~\ref{tbl1} and \ref{tbl2}, respectively.
Statistical analysis was performed for the \textit{Hinode}/SOT, \textit{SDO}/HMI, and \textit{GOES} data sets and the method of analysis and results are discussed in Sections \ref{duration} to \ref{FS}.

Statistical analysis for the hard X-ray data observed by \textit{RHESSI} was also performed.
Among the $101$ events, $27$ were simultaneously observed with \textit{RHESSI} and were shown to have greater than $50~{\rm keV}$ emissions.
Among them, $17$ WLF events ($6$ X-class and $11$ M-class flares) and $10$ NWL events ($2$ X-class and $9$ M-class flares) were observed.
An event list for the \textit{RHESSI} data is available in Table~\ref{tbl3}.
Analysis of hard X-ray data is described in Section \ref{HXR}.

\section{Statistical Data Analyses}
\label{ana}

\subsection{Flare Duration}
\label{duration}

In general, WL enhancements were found to be associated with large flares.
However, some NWL events are associated with X-class flares.
From these observational facts, it can be inferred that WLFs are associated with impulsive flares.
Using this inference, correlations with flare duration were sought.
The soft X-ray duration is easily obtained from \textit{GOES} flare information (\textit{GOES} flare start to end time).
However, some flares occurred consecutively over a short time period and it was not possible to identify the start time and/or end time from soft X-ray light curves.
It was therefore decided to use soft X-ray derivative data in order determine flare duration.
The soft X-ray derivative profile is almost the same as the hard X-ray profile from the Neupert effect \citep{Neupert1968}.
Although the Neupert effect is not always present for all flare events, this relationship was employed in this study because it is a very good index of flare duration.

\begin{figure}
\figurenum{1}
\epsscale{1.0}
\plottwo{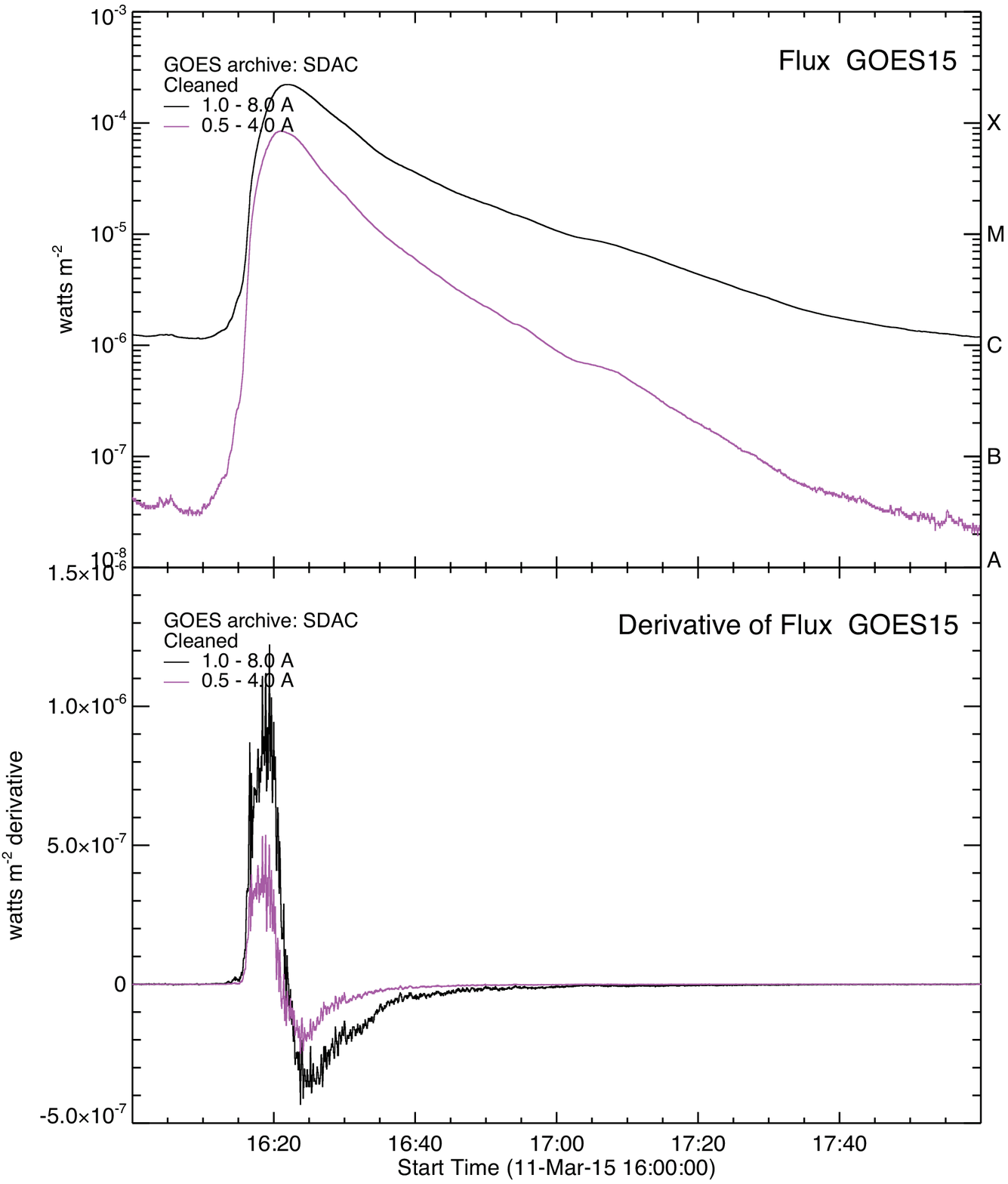} {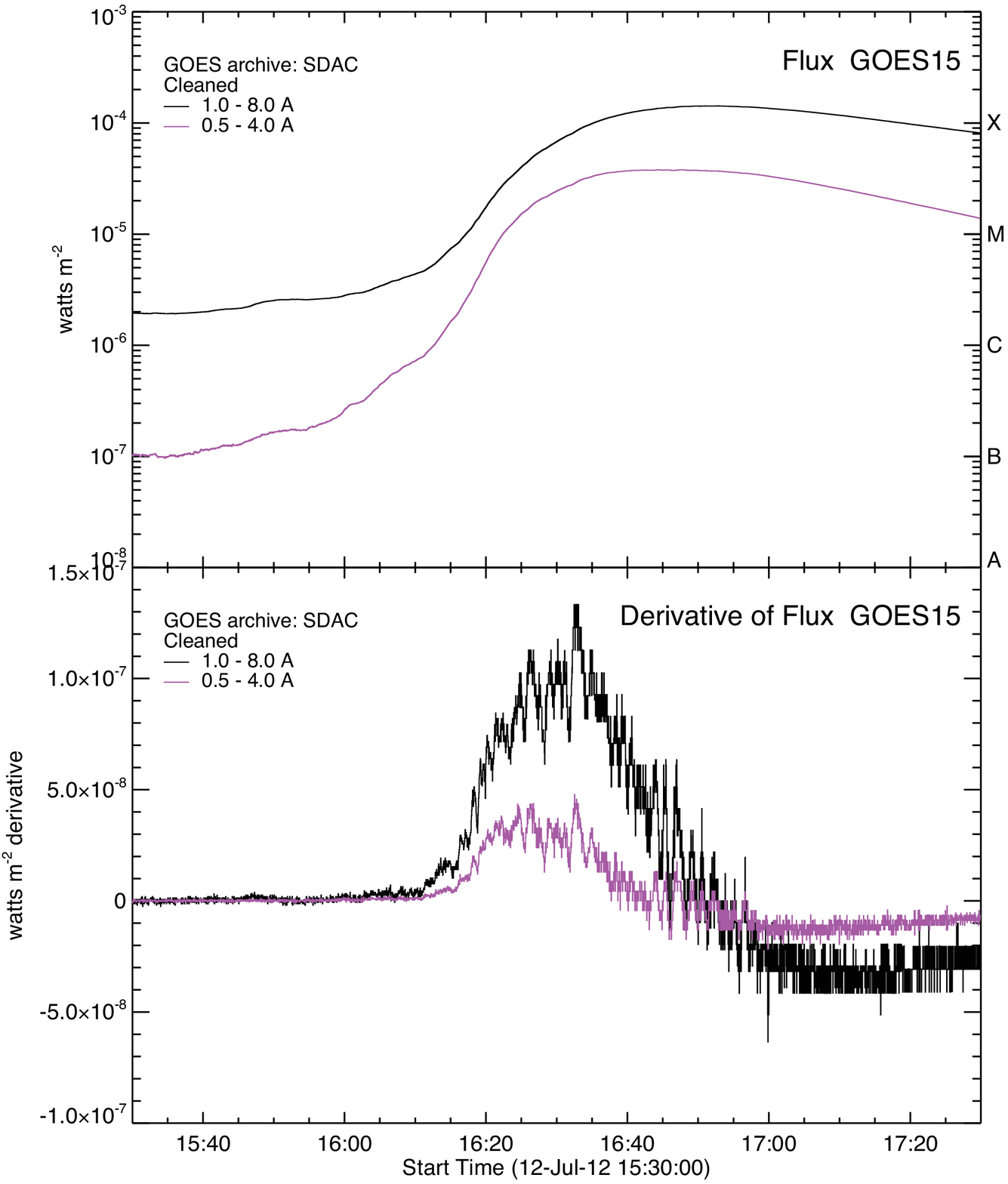} 
\caption{Sample light curves of \textit{GOES} soft X-ray flux and their derivative of flux. \textit{Left}: \textit{GOES} soft X-ray light curve (top) of the X2.1 flare on 2015 March 11 for the sample of impulsive flare and its time derivative (bottom).\textit{Right}: \textit{GOES} soft X-ray light curve (top) of the X1.4 flare on 2012 July 12 for the sample of non-impulsive flare and its time derivative (bottom). \label{fig1}}
\end{figure}

Figure~\ref{fig1} shows sample light curves of \textit{GOES} soft X-ray flux and their time derivative.
The left panels show an X2.1-class flare on 2015 March 11 as a sample of impulsive flare.
The right panels show an X1.4-class flare on 2012 July 12 as a sample of long duration event. 
Flare duration for the X1.4 flare was four times longer than that of the X2.1 flare.
From these soft X-ray derivative data, the derivative start, peak, and end time and derivative duration were obtained, as shown in Tables~\ref{tbl1} and \ref{tbl2}.
The derivative peak time is defined as the time of the peak in derivative data from the start of the \textit{GOES} flare to its peak.
The derivative start time is defined as the time when the derivative data has a continuous positive value till the derivative peak.
The derivative end time is defined as the time from the first negative value from the derivative peak time.
In some events (flares on June 13, 2012, December 05, 2014, and 15 March 2015), no \textit{GOES} data could be obtained and so no derivative data information appears in Tables~\ref{tbl1} and \ref{tbl2}.

\begin{figure}
\figurenum{2}
\epsscale{1.0}
\plottwo{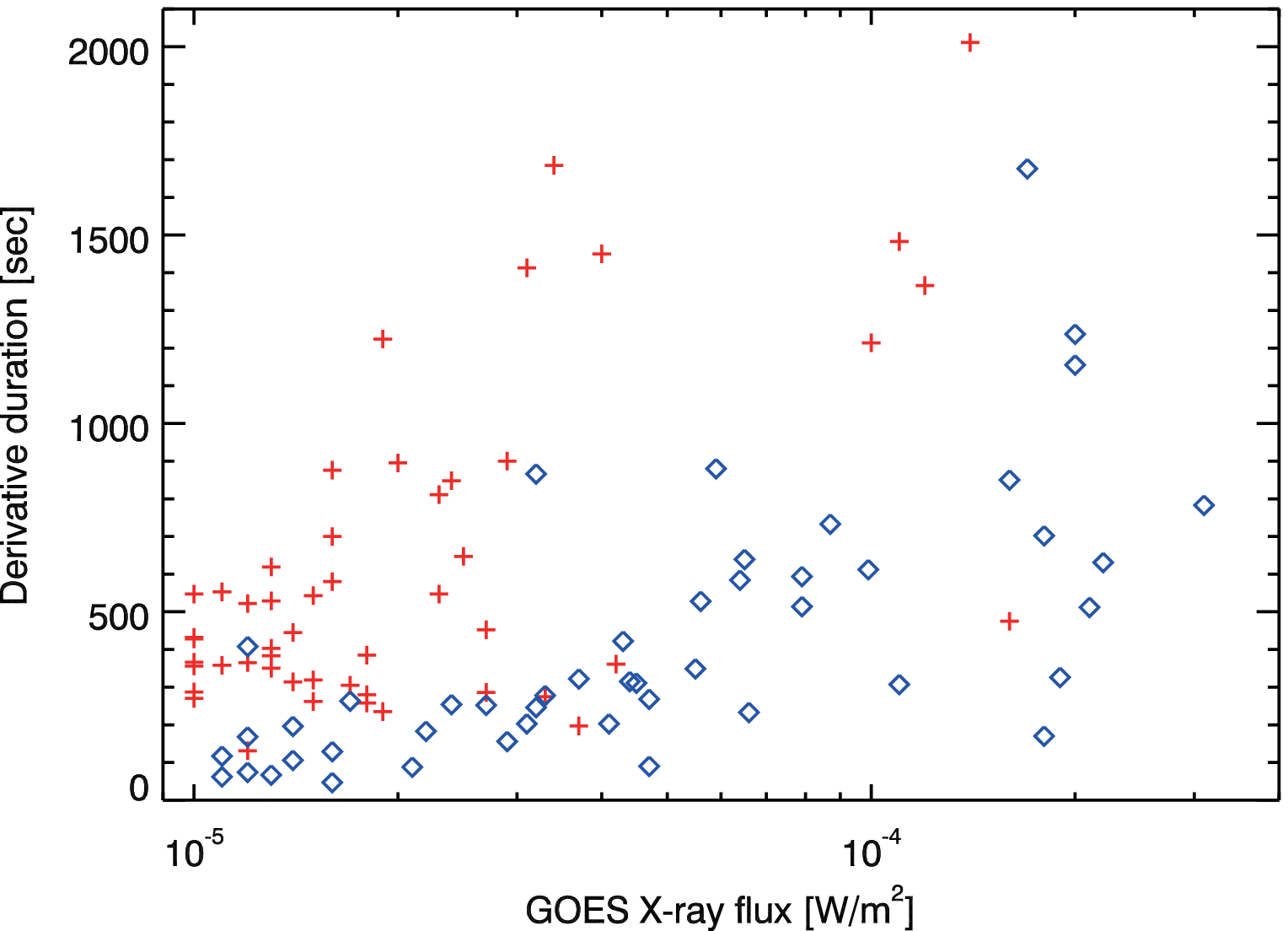} {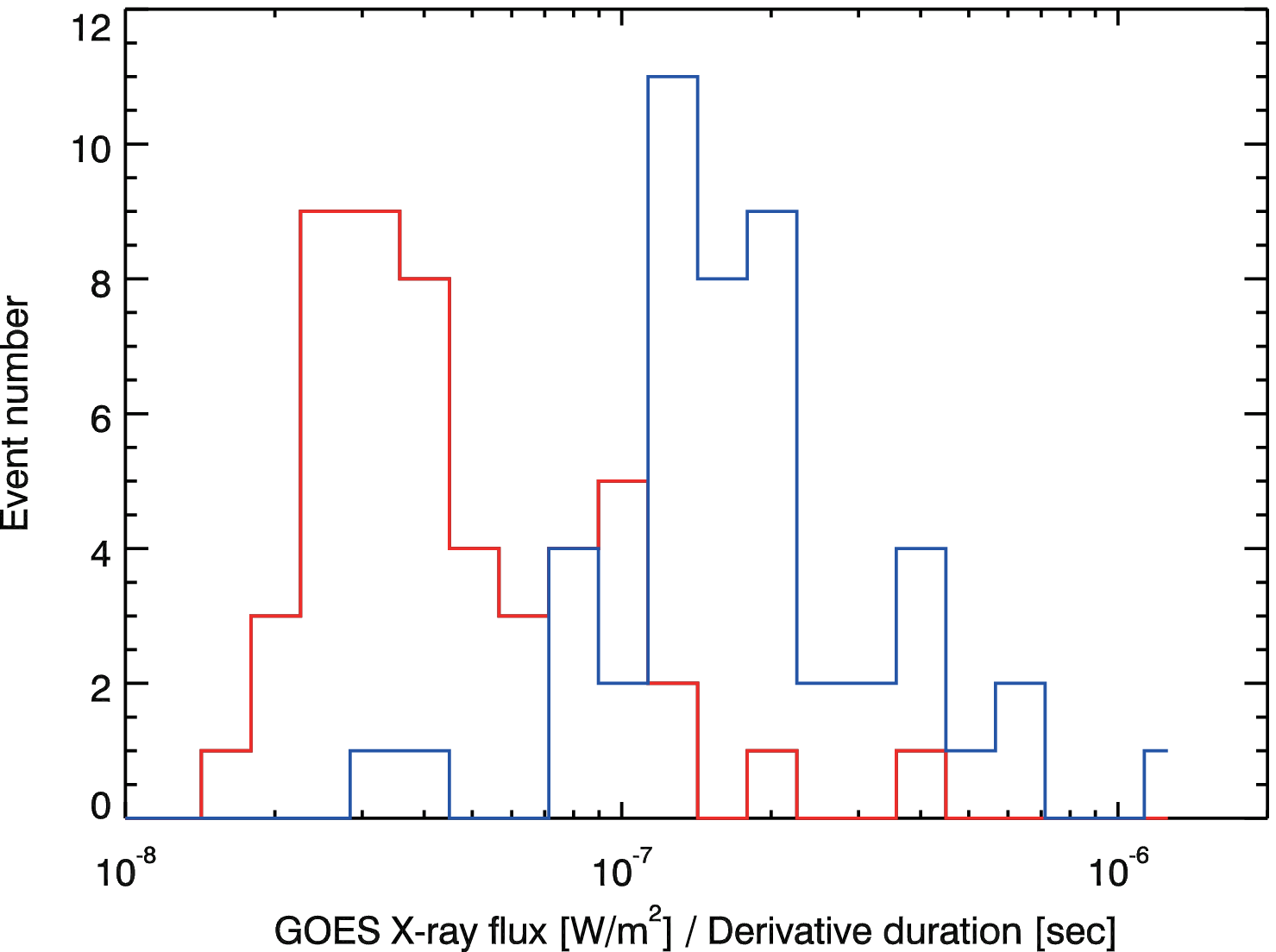} 
\caption{\textit{Left}: The relationship between the \textit{GOES} soft X-ray peak flux and the derivative duration as estimated from \textit{GOES} X-ray data. The diamond (blue) and cross (red) symbols correspond to WLF and NWL events, respectively. \textit{Right}: Histogram of event number of flare impulsivity. \label{fig2}}
\end{figure}

The relationship between the \textit{GOES} soft X-ray peak flux and the derivative duration is shown in the left-hand side panel of Figure~\ref{fig2}.
WLF events (represented by blue diamonds) show shorter duration compared with NWL events (represented by red crosses).
It appears that the flare derivative duration is roughly correlated with the \textit{GOES} soft X-ray flux.
As a result, the average of both groups could not be compared directly.
The average duration of WLF events was $419~{\rm s}$, whereas the average duration for NWL events was $619~{\rm s}$.
The average duration of NWL events is therefore $1.5$ times longer than that of WLF events.

Because there is a relationship between the \textit{GOES} X-ray peak flux and the flare derivative duration, the \textit{GOES} X-ray peak flux was divided by the derivative duration and this number was used to represent the impulsivity of the flare.
This impulsivity reflects the increase in rate of hard X-ray emission.
If there are \textit{RHESSI} data for all flare events, we don't need to use the \textit{GOES} X-ray derivative data, only the one third of flare events were observed by \textit{RHESSI} in fact.
So, we used this method.
The right-hand side panel of Figure~\ref{fig2} shows a histogram of flare impulsivity.
Figure~\ref{fig2} clearly shows two separate peaks, namely, WLF events (blue), which exhibit a shorter duration and larger flare class (impulsive flare) compared with NWL events (red).
This result indicates that the impulsivity of the flare is one of the causative factors of WL enhancement.

\subsection{Temperature and Emission Measure}
\label{TempEM}

\begin{figure}
\figurenum{3}
\epsscale{0.8}
\plotone{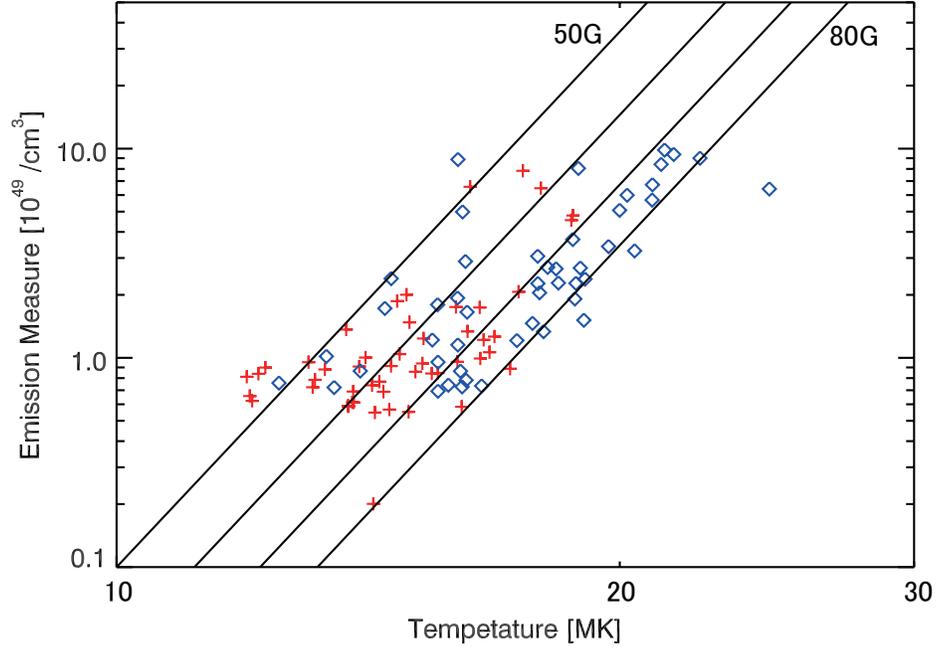} 
\caption{Scatter plot of temperature and emission measure at the derivative peak, as determined from the \textit{GOES} soft X-ray data. The diamond (blue) and cross (red) symbols correspond to WLF and NWL events, respectively. Both axes are represented using a log-scale. The relationship between emission measure and temperature based on \citet{ShibataYokoyama1999, ShibataYokoyama2002} is superposed for values of $B = 50, 60, 70$, and $80~{\rm G}$. \label{fig3}}
\end{figure}

The temperature and emission measure of each flare was calculated using the CHIANTI model.
These values were measured at the derivative end time because this time characterizes the end of the energy release from the flare.
Figure~\ref{fig3} is a scatter plot of the temperature and emission measure for all events and this graphically indicates there is a positive correlation between these two parameters.
The relationship between the emission measure and electron temperature has been reported by \citet{ShibataYokoyama1999, ShibataYokoyama2002}.
Fig.~2 in \citet{ShibataYokoyama2002} shows four theoretical curves for coronal magnetic field strengths of $5$, $15$, $50$, and  $150~{\rm G}$ at the energy-release site of the flare.
When this figure is compared with Figure~\ref{fig3} presented in this study, it can be observed that WLF and NWL events are located within the solar flare region.
It can be seen that WLF events (represented by blue diamonds) are distributed toward the right-hand side of the plot, whereas NWL events (represented by red crosses) are located on the left-hand side of the plot.
Four theoretical curves with magnetic field strength $B = 50, 60, 70, 80~{\rm G}$ are also plotted in Figure~\ref{fig3}.
Inspection of the $B = 70~{\rm G}$ curve indicates that $41$ NWL events ($82\%$ of the NWL events) are located on the left-hand side of Figure~\ref{fig3} (region of weak coronal magnetic field strength) and $31$ WLF events ($65\%$ of the WLF events) are located on the right-hand side of Figure~\ref{fig3} (region of strong coronal magnetic field strength).
The difference in the distribution of these data may be due to differences in the coronal magnetic field strength at the site of the energy release.
This result suggests that there is a tendency of the coronal magnetic field is weaker for NWL events than for WLF events.

\subsection{Distance of Flare Ribbons}
\label{ribbons}

The distance of flare ribbons is measured in order to obtain the difference between the flare formal size for WLF and NWF events.
To determine flare ribbons, the $1600~{\rm \AA}$ UV emission of the Atmospheric Imaging Assembly \citep[AIA;][]{Lemen2012} onboard \textit{SDO} was used.
From the observed two flare ribbons, the brightest point of each ribbon at the corresponding \textit{GOES} soft X-ray derivative peak time was determined and the separation between the brightest points was calculated.
Among all the events listed in Tables~\ref{tbl1} and \ref{tbl2}, flare ribbon distance for three events could not be determined due to lack of \textit{GOES} derivative data.
Furthermore, limb flares over E60 or W60 were removed from this analysis.
The \textit{SDO} data for the September 28, 2011 event were found to be missing and so these data were also not included in this analysis.
Data were corrected for the effect of parallax in order to determine flare location.

\begin{figure}
\figurenum{4}
\epsscale{1.0}
\plottwo{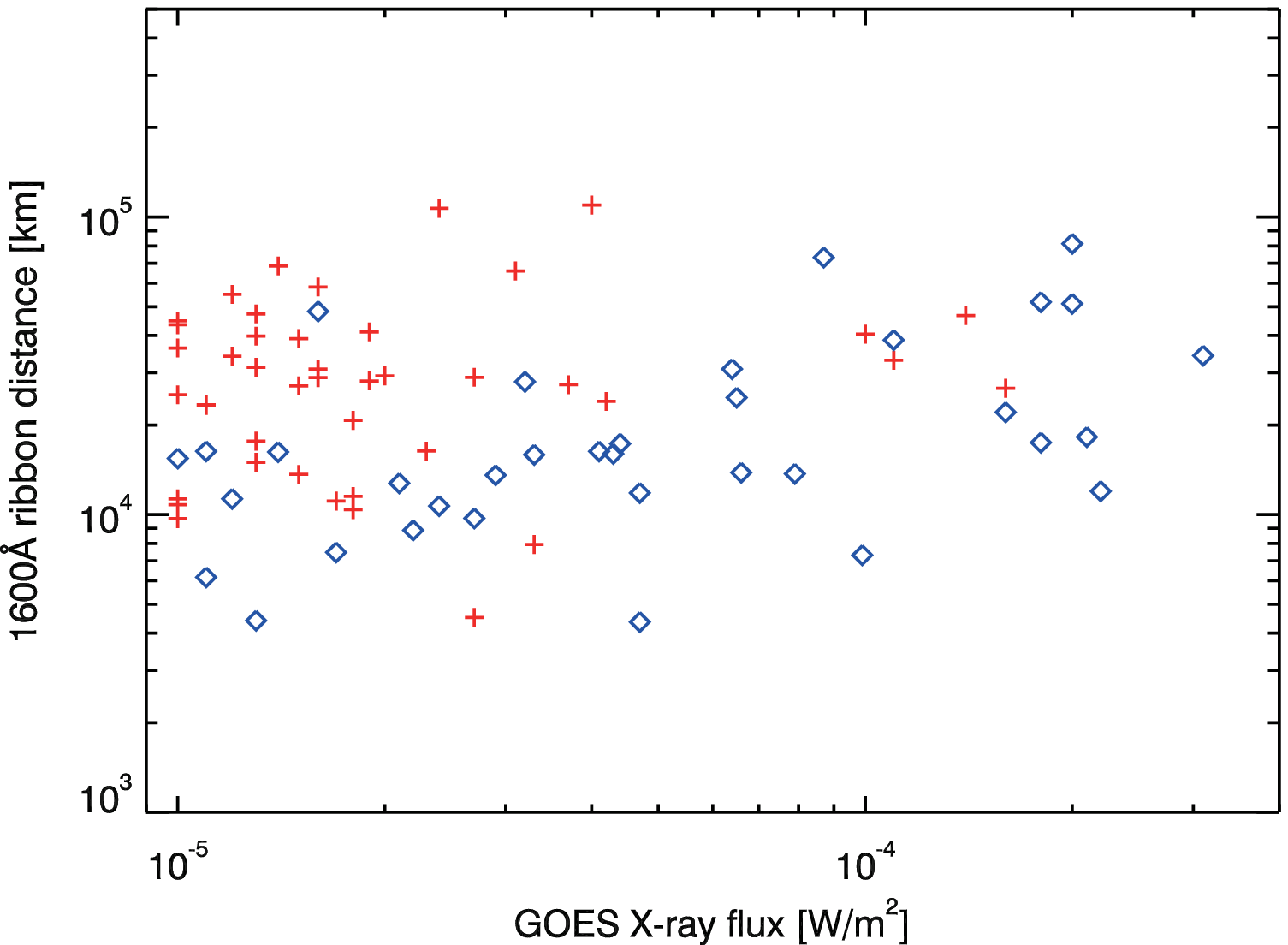} {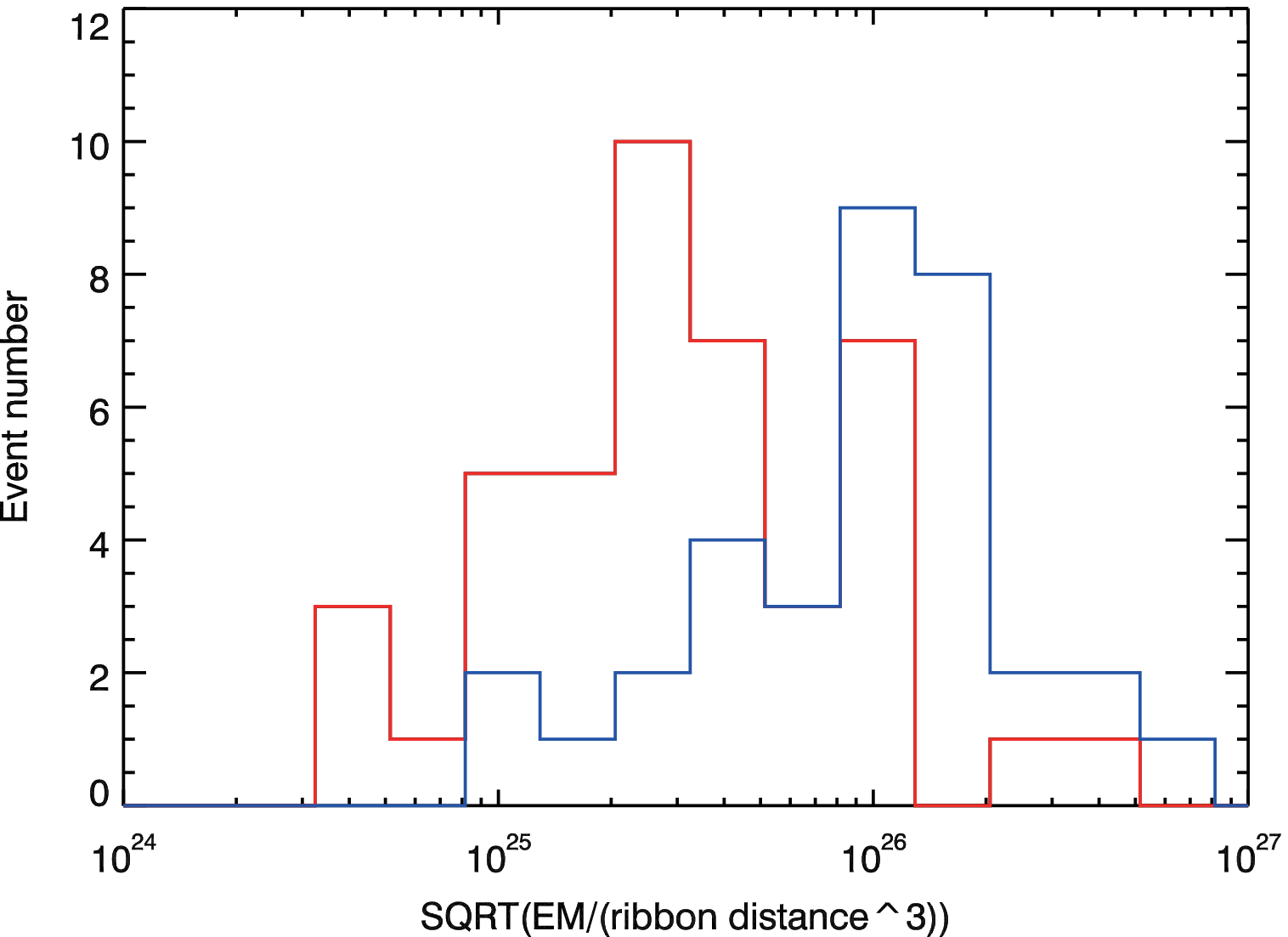} 
\caption{\textit{Left}: Relationship between \textit{GOES} soft X-ray peak flux and flare ribbon distance, as estimated from \textit{SDO}/AIA $1600~{\rm \AA}$ images (taken around derivative peak time of \textit{GOES} soft X-ray data). The diamond (blue) and cross (red) symbols correspond to WLF and NWL events, respectively. \textit{Right}: Histogram of event number for flare density. \label{fig4}}
\end{figure}

The left-hand side panel of Figure~\ref{fig4} shows the relationship between the separation of the two flare ribbons and \textit{GOES} soft X-ray class.
It appears that there is a rough correlation between the ribbon distance and soft X-ray class, i.e., intense flares have a relatively large size.
However, the ribbon distance of NWL events is located in the upper part of the left-hand side panel of Figure~\ref{fig4} compared with that of WLF events.
The average separation for WLF events was determined as $2.2 \times 10^3 ~{\rm km}$ and that for NWL events as $3.3 \times 10^3~{\rm km}$.
The separation of WLF events is therefore significantly less than that of NWL events.
This result indicates that the flare formal size of WLF events is relatively more compact than that of NWL events.

This flare ribbon distance is an index of flare size.
When the emission measure is divided by flare volume (distance$^3$) and the square-root taken, an index of flare loop density can be obtained.
The right-hand side panel of Figure~\ref{fig4} shows the histogram of event number for flare density.
In this figure, it is just possible to see two peaks for WLF events (blue) and for NWL events (red).
This result indicates that the plasma density of the flare is one of the causative factors of WL enhancement.

\subsection{Field Strength Under the Flare Ribbons}
\label{FS}

From Section  \ref{TempEM}, it was suggested that there is a relationship between WLF and NWL events and the coronal magnetic field strength.
However, no coronal magnetic field strength data exist in the observational data used in this study.
Instead, the photospheric magnetic field strength under the flare ribbons as obtained in Section \ref{ribbons} is employed.

The \textit{SDO}/HMI field strength data under the $1600~{\rm \AA}$ flare ribbons is therefore used in this analysis.
For the two flare ribbons associated with each flare, positive field strength is taken from one flare ribbon and negative field strength taken from the other ribbon.
Most of the event, these flare ribbons were located on plage region, not umbrae.
We calculated the average field strength under the $1600~{\rm \AA}$ flare ribbons.

\begin{figure}
\figurenum{5}
\epsscale{0.8}
\plotone{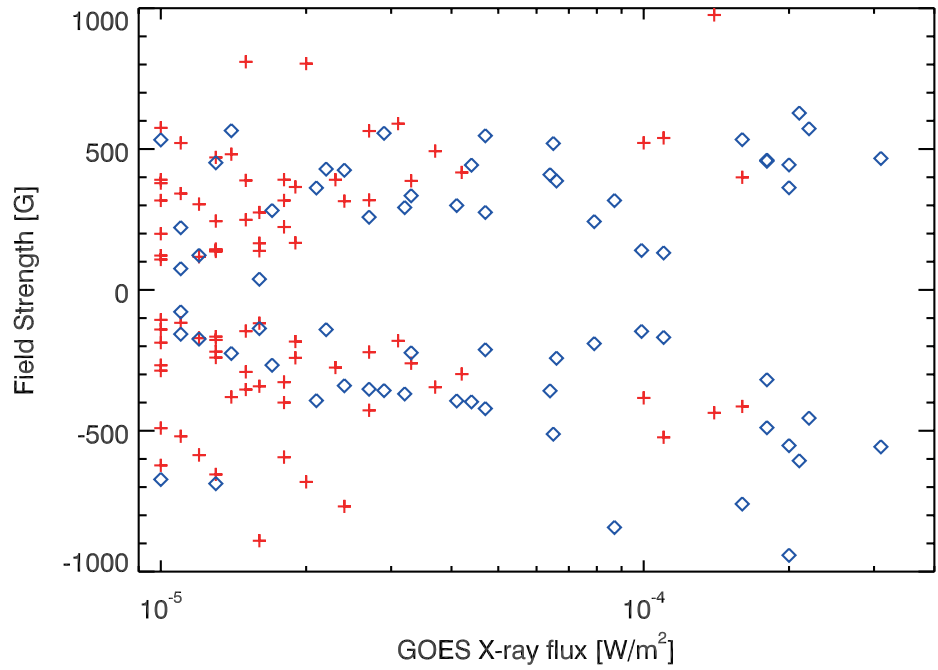} 
\caption{Relationship between \textit{GOES} soft X-ray peak flux and field strength under the $1600~{\rm \AA}$  flare ribbons, as estimated from \textit{SDO}/HMI data (taken around derivative peak time of \textit{GOES} soft X-ray data). The diamond (blue) and cross (red) symbols correspond to WLF and NWL events, respectively. \label{fig5}}
\end{figure}

Figure~\ref{fig5} shows the relationship between the \textit{GOES} soft X-ray peak flux and the field strength under the $1600~{\rm \AA}$ flare ribbons as estimated from \textit{SDO}/HMI data (taken from around the derivative peak time of \textit{GOES} soft X-ray data).
Unfortunately there was a high degree of scatter in the data and no relationship could be determined between these two parameters.

\subsection{Hard X-ray Data Analysis}
\label{HXR}

As mentioned in Section \ref{event}, a statistical analysis of hard X-ray data as observed by \textit{RHESSI} was performed. Among the $101$ \textit{Hinode}/SOT events, $46$ flare events were simultaneously observed with \textit{RHESSI}.
Among the $46$ flare events, $21$ events were associated with the WLF events and $25$ events were the result of NWL events.
Among them, $17$ WLF events ($6$ X-class and $11$ M-class flares) and $10$ NWL events ($2$ X-class and $9$ M-class flares) have emission greater than $50~{\rm keV}$.
Physical parameters for these events are given in Table~\ref{tbl3}.
Based on these events, hard X-ray photon counts and spectra could be analyzed.

\subsubsection{Maximum Photon Counts of Hard X-rays}
\label{HXRpeak}

\begin{figure}
\figurenum{6}
\epsscale{0.8}
\plotone{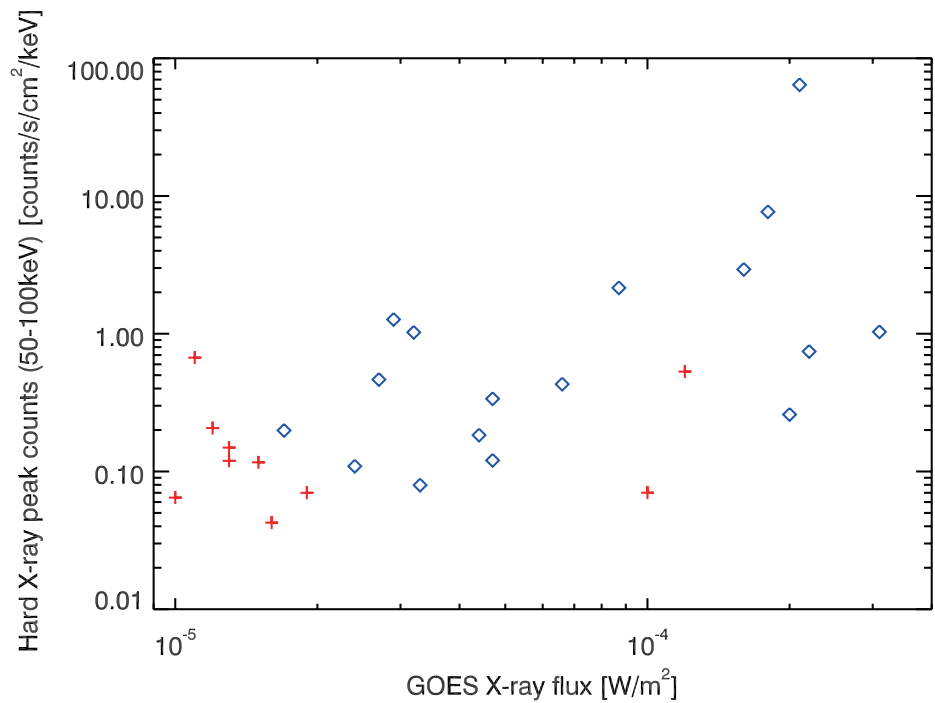} 
\caption{Relationship between \textit{GOES} soft X-ray peak flux and $50-100~{\rm keV}$ hard X-ray peak count obtained from \textit{RHESSI} data. The diamond (blue) and cross (red) symbols correspond to WLF and NWL events, respectively. \label{fig6}}
\end{figure}

The maximum photon count of $50-100~{\rm keV}$ hard X-ray photons during the flares was investigated.
The photon count of each flare is given in Table~\ref{tbl3}.
A scatter plot of the photon count and \textit{GOES} soft X-ray flux is shown in Figure~\ref{fig6}.
Consideration of Figure~\ref{fig6} shows that hard X-ray photon count is roughly correlated with the \textit{GOES} soft X-ray flux.
The average photon count of the hard X-rays was calculated and found to be approximately $4.6~{\rm counts/s/cm^2/keV}$ for WLF events and $0.2~{\rm counts/s/cm^2/keV}$ for NWL events.
Although the standard deviation is larger than the above mentioned values, the maximum photon count of WLF events is significantly larger than that of NWL events.
This is directly related to the fact that the \textit{GOES} soft X-ray flux for WLF events is larger than that for NWL events.

\subsubsection{Hard X-ray Spectra and Non-thermal Energy}
\label{HXRene}

\begin{figure}
\figurenum{7}
\epsscale{0.8}
\plotone{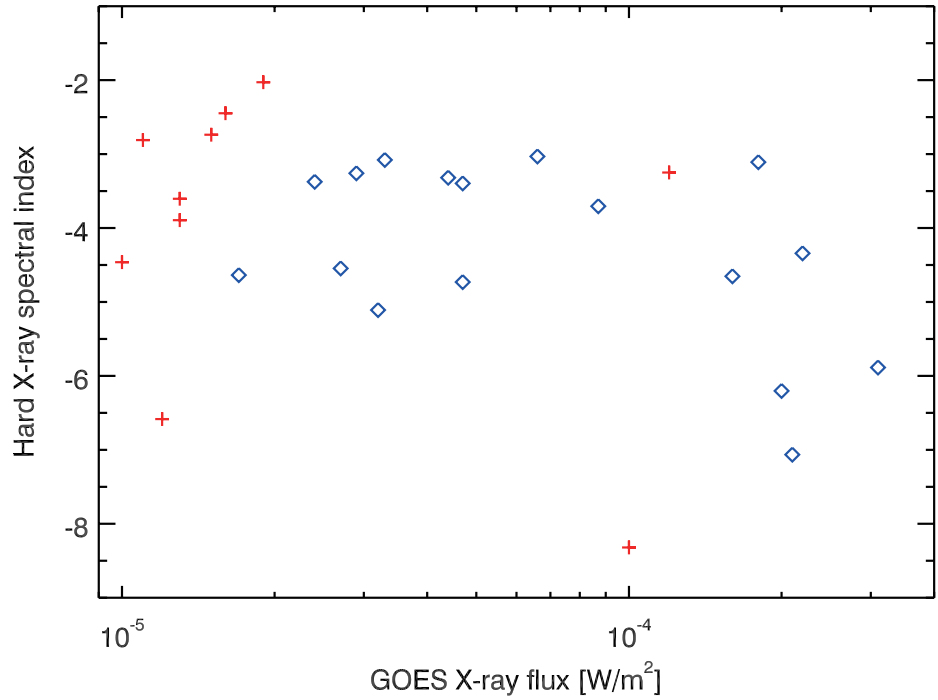} 
\caption{Relationship between \textit{GOES} soft X-ray peak flux and spectral index of hard X-ray, as estimated from \textit{RHESSI} data. The diamond (blue) and cross (red) symbols correspond to WLF and NWL events, respectively. \label{fig7}}
\end{figure}

Spectral fitting was performed for $27$ events using a single power law to describe each hard X-ray spectrum in the region of the hard X-ray peak time and in the range of $30-100~{\rm keV}$.
Indices for the power laws obtained are given in Table~\ref{tbl3} and the correlation of spectral index and \textit{GOES} soft X-ray flux are given in Figure~\ref{fig7}.
The average power-law indices are $-4.3$ for WLF events and $-4.0$ for NWL events.
Because the standard deviation is smaller than $-1.0$, there is no significant difference in the power-law indices.

\begin{figure}
\figurenum{8}
\epsscale{0.8}
\plotone{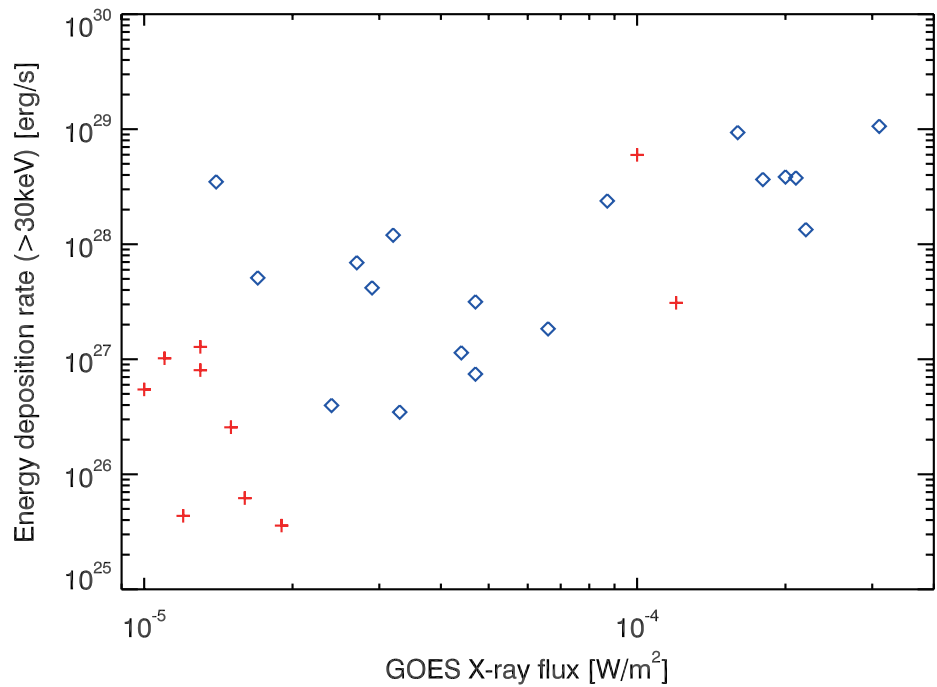} 
\caption{Relationship between \textit{GOES} soft X-ray peak flux and energy deposition rate of  $>30~{\rm keV}$ emission, as estimated from \textit{RHESSI} data. The diamond (blue) and cross (red) symbols correspond to WLF and NWL events, respectively. \label{fig8}}
\end{figure}

The deposition rate of non-thermal energy was then calculated assuming a thick target model with a low-energy cutoff of $30~{\rm keV}$ \citep{Brown1971}.
This was done using the same method as described in \citep{Watanabe2010}.
The calculated deposition rate of the non-thermal energy and \textit{GOES} soft X-ray flux are given in Table~\ref{tbl3}.
The corresponding scatter plot is found in Figure~\ref{fig8}.
There is one NWL event in Figure~\ref{fig8} with very high energy deposition.
That is the X1.0-class flare on 2014 October 25, and that is due to the very soft power index for this event.
The average deposition rate of non-thermal energy was found to be $2.34 \pm 3.05 \times 10^{28}~{\rm erg/s}$ for WLF events and $6.68 \pm 1.77 \times 10^{27}~{\rm erg/s}$ for NWL events.
The deposition rate of WLF events is significantly larger than that of NWL events and this is because the \textit{GOES} soft X-ray flux for WLF events is larger than that for NWL events.

\begin{figure}
\figurenum{9}
\epsscale{0.8}
\plotone{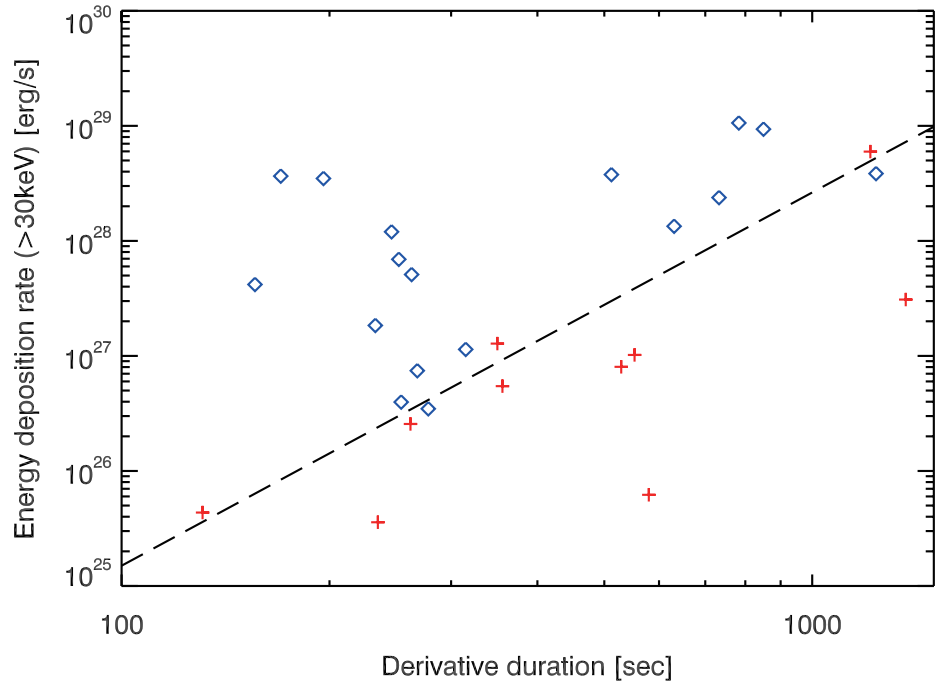} 
\caption{Relationship between the derivative duration estimated from \textit{GOES} X-ray data and energy deposition rate of $>30~{\rm keV}$ emission, as estimated from \textit{RHESSI} data. The diamond (blue) and cross (red) symbols correspond to WLF and NWL events, respectively. A dashed line separating the two populations which would help to "guide the eye". \label{fig9}}
\end{figure}

The X-axis from the \textit{GOES} soft X-ray flux is then plotted as the derivative duration, and the results are shown in Figure~\ref{fig9}.
In Figure~\ref{fig9}, WLF events are located on the upper left-hand side of the dashed line and NWL evens are located on the lower right-hand side of the dashed line.
It is important to note that these data occupy separate regimes in Figure~\ref{fig9}.
This result suggests that the injection rate of non-thermal electrons is one of the causative factors of WL enhancement.

\section{Discussion}
\label{discussion}

The results described in the previous sections can be summarized as follows:
\begin{enumerate}
\renewcommand{\labelenumi}{(\arabic{enumi})}
\item The derivative duration of \textit{GOES} soft X-rays during WLF events is relatively short.
\item WLF events are characterized by stronger magnetic fields at the energy-release site compared with NWL events from the relationship between the temperature and emission measure.
\item The separation between two ribbons is shorter for WLF events.
\item No significant difference exists in field strength under the flare ribbons.
\item The hard X-ray photon count in the energy range of $50-100~{\rm keV}$ and the deposition rate of non-thermal energy ($>30~{\rm keV}$) are correlated with the \textit{GOES} soft X-ray flux.
\item No significant difference exists between the power-law indices of hard X-ray spectra for WLF and NWL events.
\item WLF events are characterized by larger deposition rates of non-thermal energy compared with NWL events. WLF events are characterized by a shorter derivative duration for a given deposition rate.
\end{enumerate}

From results (1) and (3), WLF events appear to be small in terms of spatial extent and have a relatively short duration.
WLF events are therefore short-lived and exhibit a rapid enhancement phase, whereas NWL events show a gradual enhancement phase.
The short separation of flare ribbons implies that the magnetic field structure related to WLF events is compact.
During the evolution of a flare, the separation between flare ribbons increases as a function of time, which is consistent with the short duration.
Result (3) suggests that the magnetic loop related to WLF events occurs at low altitudes.
In general, the magnetic field is stronger at low altitudes than at high altitudes.
The energy release of WLF events therefore appears to occur in the lower corona wherein the magnetic field is relatively strong.

Result (2) suggests the existence of strong magnetic fields at the energy-release site for WLF events.
The scatter plot of the emission measure and electron temperature reported in Fig.~2 of \citet{ShibataYokoyama2002} shows four theoretical curves for coronal magnetic field strengths of $5$, $15$, $50$, and $150~{\rm G}$ at the energy-release site of the flare.
The scatter plot of emission measure and electron temperature reported in this investigation (Fig.~2) shows that WLF events are distributed toward the right-hand side of plot, whereas NWL events are located on the left-hand side of the plot.
The difference in the distribution of these data may be due to differences in coronal magnetic field strength at the energy-release site.
The magnetic field is weaker for NWL events than for WLF events.
This finding is consistent with result (3) as the magnetic field is stronger at lower altitudes in the corona.

An attempt was made to check the field strength of the photosphere.
However, due to lack of field strength data at the energy-release site, an alternative analysis was undertaken, as described in Section \ref{FS}.
No significant difference was observed in the field strength under the flare ribbons.

Results (5) to (7) relate to the hard X-ray observations.
As described in Section \ref{intro}, WL enhancement is correlated with hard X-ray sources in time, location, and energy.
Contrary to the expectations of this study, no significant difference was observed in the power-law indices of WLF and NWL events.
It therefore follows that the energy distribution of accelerated electrons is similar in both WLF and NWL events.
Only very high-energy ($\gg100~{\rm keV}$) electrons can reach the photosphere \citep{AboudarhamHenoux1986, Neidig1989}.
Results presented in this work do not show that the fraction of electrons with such high energies is larger in WLF events compared with NWL events.
This result does not support the model whereby electrons penetrate directly into the photosphere and emit WL.
However, as the conclusions presented here are the result of a statistical analysis, it is possible that some WLF events may result from this process.

The results presented in this investigation indicate that in order to enhance WL emissions, a large number of accelerated electrons must precipitate within a short period, thereby leading to very rapid heating of the atmosphere.
Result (7) suggests that rapid heating is important and that a threshold exists in terms of the injection rate of non-thermal electrons required to generate WL emission.

Through this statistical study, it has been shown that WLF events are characterized by stronger magnetic fields at the energy-release site compared with NWL events.
This strong magnetic field may be a factor in the enhancement of WL emission.
It is important to consider how a strong magnetic field may be related to WLF events.
Two models explain WL emission and indicate that the energy-release region is characterized by a strong magnetic field.
In the solar flare model based on magnetic reconnection, the energy-release rate increases if the energy-release region has a strong magnetic field and electrons are accelerated.
Hard X-ray sources appear at the region wherein the magnetic field is strongest along the flare ribbon \citep{Asai2002}.
Large numbers of electrons are accelerated and this leads to the enhancement of WL emission.
The strong magnetic field may also be considered in terms of the trapping efficiency of accelerated electrons in the flare loop.
If the top part of the loop is characterized by a stronger magnetic field than that of the foot-point region, the magnetic mirror ratio between the top and the foot-point becomes small and the loss-cone angle is large.
Because of this effect, a larger number of accelerated electrons can precipitate into the foot-point region within a short period of time.
The trapping efficiency of the loop is therefore lower and most of the accelerated electrons precipitate directly into the foot-point region.
This scenario is consistent with results obtained in this study.

Investigating the magnetic structure of a flare region could prove interesting and may reveal how and why the magnetic field strength at the energy-release site is related to the generation mechanism of WLF events.

\section{Conclusion}
\label{conclusion}

A statistical study has been presented comprising $101$ solar flares observed using the visible continuum filter of \textit{Hinode}/SOT and \textit{SDO}/HMI for \textit{GOES} M- and X-class flares occurring during the period from January 2011 to February 2016.
Of these $101$ events, $49$ WLF events and $52$ NWL events were identified on the basis of the existence of an enhancement of the visible continuum images obtained using \textit{Hinode}/SOT and \textit{SDO}/HMI.
The WLF events are characterized by short duration, a high temperature in the \textit{GOES} soft X-ray data, and short distance between two flare ribbons in the \textit{SDO}/AIA $1600~{\rm \AA}$ observations.
No significant difference was observed between WLF and NWL events in terms of power-law indices, flux of non-thermal photons ($50-100~{\rm keV}$), or in the deposition rate of the non-thermal energy.
However, a clear relationship between the injection rate of non-thermal energy and derivative duration was observed.
These results indicate that during WLF events, accelerated electrons precipitate in a short time period, thereby leading to rapid heating of the atmosphere.
The similarity in the power-law index of hard X-ray spectra, as well as the similar deposition rate of non-thermal energy, does not indicate that the fraction of electrons with very high energies ($\gg100~{\rm keV}$) is larger in WLF events compared with NWL events.
Relatively low-energy ($<100~{\rm keV}$) electrons appear to contribute to the enhancement of WL.
This finding is consistent with studies detailing the energy budget between non-thermal electrons and WL emission \citep{Watanabe2010}.

The statistical analysis presented here suggests that the non-thermal energy deposition rate and the magnetic field strength at the energy-release site are significant in WL emission in solar flares.
In future work, investigating the magnetic field structure around the energy-release site would be interesting to describe more fully the environment of non-thermal electrons.
The physical relationship between the magnetic field structure and the energy of electrons should therefore be the subject of a detailed investigation.
Such a study would be expected to provide insight into not only the process of WL emission but also into the acceleration of electrons in solar flares.

It is useful to derive the color temperature of WLF when its energy source is studied \citep{Watanabe2013, Kerr2014}.
However, we didn't derive it in this paper because we focused on the difference of WLF and NWL events.
In future, we would like to perform this kind of detailed WLF analyses which would provide unique constraints for radiative-hydrodynamic flare models and might reveal lower atmospheric heating differences in impulsive flares and non-impulsive flares.
Moreover, \citet{Kowalski2013} showd how the NUV and optical continuum spectral properties of M dwarf flares vary from impulsive flare events to gradual flare events, where the type of flare (impulsive vs. gradual) is determined from a similar quantity as the "impulsivity".
It would also be interesting to discuss how their results for M dwarf WLFs connect with our study.

\bigskip

\textit{Hinode} is a Japanese mission developed and launched by ISAS/JAXA, with NAOJ as domestic partner and NASA and STFC (UK) as international partners. It is operated by these agencies in co-operation with ESA and NSC (Norway).
This study was supported by the JSPS KAKENHI Grant Numbers JP15K17622, JP16H01187.
This work was performed by the joint research program of the Institute for Space-Earth Environmental Research (ISEE), Nagoya University.




\floattable
\begin{deluxetable}{ccccccrccccc}
\tablecaption{WLF event list and physical parameters from \textit{GOES} \& \textit{SDO} data. \label{tbl1}}
\rotate
\tablecolumns{12}
\tablenum{1}
\tablewidth{0pt}
\tablehead{
\colhead{} & \colhead{} & \colhead{} & \colhead{} & \colhead{GOES} & \colhead{} & \colhead{derivative} & \multicolumn{2}{c}{@derivative end} & \colhead{$1600{\rm \AA}$ ribbon} & \multicolumn{2}{c}{HMI field strength} \\
\cline{8-9}
\colhead{GOES flare start} & \multicolumn{3}{c}{GOES derivative} & \colhead{X-ray} & \colhead{Sunspot} & \colhead{duration} & \colhead{Temperature} & \colhead{EM} & \colhead{distance} & \multicolumn{2}{c}{@$1600{\rm \AA}$ ribbon} \\
\cline{2-4}\cline{11-12}
\colhead{\scriptsize{YYYY/MM/DD hh:mm}} & \colhead{start} & \colhead{peak} & \colhead{end} & \colhead{class} & \colhead{location} & \colhead{[sec]} & \colhead{[MK]} & \colhead{[$10^{49}/{\rm cm}^3$]} & \colhead{[$\times 10^3 {\rm km}$]} & \colhead{positive [G]} & \colhead{negative [G]}
}
\startdata
2011/02/15 01:44 & 2011/02/15 01:45 & 2011/02/15 01:52 & 2011/02/15 01:56 & X2.2 & S20W10 & 631 & 21.3 & 9.85 & 12.0 & 572.4 & -455.2\\
2011/11/03 20:16 & 2011/11/03 20:17 & 2011/11/03 20:21 & 2011/11/03 20:22 & X1.9 & N21E64 & 326 & 20.9 & 5.68 &  &  & \\
2012/01/27 17:37 & 2012/01/27 18:05 & 2012/01/27 18:26 & 2012/01/27 18:33 & X1.7 & N33W85 & 1676 & 16.0 & 8.88 &  &  & \\
2012/07/06 23:01 & 2012/07/06 23:03 & 2012/07/06 23:06 & 2012/07/06 23:08 & X1.1 & S13W59 & 307 & 20.0 & 5.07 & 38.6 & 131.6 & -168.4\\
2012/10/23 03:13 & 2012/10/23 03:14 & 2012/10/23 03:15 & 2012/10/23 03:16 & X1.8 & S13E58 & 170 & 24.6 & 6.42 & 17.5 & 460.9 & -488.9\\
2014/10/22 14:02 & 2014/10/22 14:03 & 2014/10/22 14:15 & 2014/10/22 14:17 & X1.6 & S14E13 & 850 & 20.2 & 5.99 & 22.1 & 533.8 & -759.8\\
2014/10/24 20:50 & 2014/10/24 21:06 & 2014/10/24 21:15 & 2014/10/24 21:19 & X3.1 & S16W21 & 783 & 22.3 & 9.00 & 34.3 & 466.9 & -557.0\\
2014/10/26 10:04 & 2014/10/26 10:35 & 2014/10/26 10:44 & 2014/10/26 10:54 & X2.0 & S18W40 & 1155 & 21.2 & 8.41 & 51.1 & 443.9 & -942.0\\
2014/10/27 14:12 & 2014/10/27 14:10 & 2014/10/27 14:26 & 2014/10/27 14:31 & X2.0 & S17W52 & 1237 & 20.9 & 6.72 & 81.4 & 363.0 & -552.5\\
2014/12/20 00:11 & 2014/12/20 00:14 & 2014/12/20 00:21 & 2014/12/20 00:25 & X1.8 & S21W24 & 702 & 18.9 & 8.05 & 51.8 & 457.0 & -318.8\\
2015/03/11 16:11 & 2015/03/11 16:13 & 2015/03/11 16:19 & 2015/03/11 16:21 & X2.1 & S17E22 & 512 & 21.5 & 9.38 & 18.2 & 627.8 & -606.4\\
2011/02/14 17:20 & 2011/02/14 17:23 & 2011/02/14 17:24 & 2011/02/14 17:26 & M2.2 & S20W04 & 183 & 15.4 & 1.22 & 8.9 & 429.2 & -140.8\\
2011/02/18 09:55 & 2011/02/18 10:07 & 2011/02/18 10:10 & 2011/02/18 10:11 & M6.6 & S21W55 & 233 & 20.4 & 3.25 & 13.8 & 386.9 & -242.4\\
2011/02/18 12:59 & 2011/02/18 13:00 & 2011/02/18 13:02 & 2011/02/18 13:03 & M1.4 & S21W55 & 196 & 16.2 & 0.78 & 16.2 & 565.5 & -225.3\\
2011/09/28 13:24 & 2011/09/28 13:25 & 2011/09/28 13:26 & 2011/09/28 13:28 & M1.2 & N11E00 & 168 & 15.6 & 0.69 &  &  & \\
2011/11/02 21:52 & 2011/11/02 21:53 & 2011/11/02 21:57 & 2011/11/02 22:00 & M4.3 & N20E77 & 422 & 17.9 & 2.05 &  &  & \\
2011/12/31 13:09 & 2011/12/31 13:11 & 2011/12/31 13:13 & 2011/12/31 13:15 & M2.4 & S25E46 & 254 & 17.4 & 1.21 & 10.7 & 425.1 & -339.9\\
2012/03/06 12:23 & 2012/03/06 12:39 & 2012/03/06 12:40 & 2012/03/06 12:41 & M2.1 & N21E40 & 88 & 16.0 & 1.15 & 12.8 & 362.1 & -392.8\\
2012/03/06 21:04 & 2012/03/06 21:09 & 2012/03/06 21:10 & 2012/03/06 21:11 & M1.3 & N16E30 & 67 & 16.1 & 0.73 & 4.4 & 451.4 & -687.6\\
2012/05/09 12:21 & 2012/05/09 12:27 & 2012/05/09 12:29 & 2012/05/09 12:32 & M4.7 & N13E31 & 268 & 18.4 & 2.28 & 11.8 & 275.4 & -212.2\\
2012/05/09 21:01 & 2012/05/09 21:01 & 2012/05/09 21:03 & 2012/05/09 21:05 & M4.1 & N12E26 & 203 & 18.8 & 1.91 & 16.3 & 299.9 & -393.8\\
2012/05/10 20:20 & 2012/05/10 20:21 & 2012/05/10 20:25 & 2012/05/10 20:26 & M1.7 & N12E12 & 263 & 15.6 & 0.95 & 7.5 & 282.0 & -267.2\\
2012/07/05 03:25 & 2012/07/05 03:34 & 2012/07/05 03:35 & 2012/07/05 03:36 & M4.7 & S18W29 & 90 & 18.8 & 2.27 & 4.4 & 547.5 & -421.1\\
2012/07/05 21:37 & 2012/07/05 21:44 & 2012/07/05 21:45 & 2012/07/05 21:45 & M1.6 & S12W46 & 47 & 13.4 & 1.02 & 48.2 & 38.3 & -136.8\\
2012/07/06 01:37 & 2012/07/06 01:37 & 2012/07/06 01:38 & 2012/07/06 01:40 & M2.9 & S18W41 & 156 & 17.7 & 1.46 & 13.6 & 556.3 & -357.1\\
\enddata
\end{deluxetable}

\floattable
\begin{deluxetable}{ccccccrccccc}
\tablecaption{WLF event list and physical parameters from \textit{GOES} \& \textit{SDO} data. (cont.)}
\rotate
\tablecolumns{12}
\tablenum{1}
\tablewidth{0pt}
\tablehead{
\colhead{} & \colhead{} & \colhead{} & \colhead{} & \colhead{GOES} & \colhead{} & \colhead{derivative} & \multicolumn{2}{c}{@derivative end} & \colhead{$1600{\rm \AA}$ ribbon} & \multicolumn{2}{c}{HMI field strength} \\
\cline{8-9}
\colhead{GOES flare start} & \multicolumn{3}{c}{GOES derivative} & \colhead{X-ray} & \colhead{Sunspot} & \colhead{duration} & \colhead{Temperature} & \colhead{EM} & \colhead{distance} & \multicolumn{2}{c}{@$1600{\rm \AA}$ ribbon} \\
\cline{2-4}\cline{11-12}
\colhead{\scriptsize{YYYY/MM/DD hh:mm}} & \colhead{start} & \colhead{peak} & \colhead{end} & \colhead{class} & \colhead{location} & \colhead{[sec]} & \colhead{[MK]} & \colhead{[$10^{49}/{\rm cm}^3$]} & \colhead{[$\times 10^3 {\rm km}$]} & \colhead{positive [G]} & \colhead{negative [G]}
}
\startdata
2012/07/08 12:05 & 2012/07/08 12:08 & 2012/07/08 12:09 & 2012/07/08 12:10 & M1.4 & S21W69 & 106 & 16.5 & 0.73 &  &  & \\
2013/06/07 22:11 & 2013/06/07 22:33 & 2013/06/07 22:43 & 2013/06/07 22:48 & M5.9 & S32W89 & 880 & 18.9 & 2.69 &  &  & \\
2013/10/26 19:24 & 2013/10/26 19:23 & 2013/10/26 19:24 & 2013/10/26 19:27 & M3.1 & S09E81 & 203 & 14.5 & 1.72 &  &  & \\
2013/10/28 15:07 & 2013/10/28 15:09 & 2013/10/28 15:10 & 2013/10/28 15:14 & M4.4 & S06E28 & 315 & 14.6 & 2.39 & 17.3 & 443.2 & -397.2\\
2013/12/22 14:45 & 2013/12/22 15:07 & 2013/12/22 15:10 & 2013/12/22 15:12 & M3.3 & S19W56 & 278 & 15.6 & 1.80 & 15.9 & 334.4 & -223.1\\
2013/12/23 08:59 & 2013/12/23 09:04 & 2013/12/23 09:06 & 2013/12/23 09:06 & M1.6 & S17W63 & 129 & 16.1 & 0.86 &  &  & \\
2013/12/31 21:45 & 2013/12/31 21:47 & 2013/12/31 21:52 & 2013/12/31 21:57 & M6.4 & S16W35 & 584 & 17.9 & 3.06 & 30.9 & 409.1 & -358.2\\
2014/01/01 18:40 & 2014/01/01 18:41 & 2014/01/01 18:47 & 2014/01/01 18:51 & M9.9 & S14W47 & 612 & 16.1 & 4.99 & 7.3 & 140.5 & -147.2\\
2014/03/13 19:03 & 2014/03/13 19:10 & 2014/03/13 19:12 & 2014/03/13 19:17 & M1.2 & N15W87 & 408 & 13.5 & 0.72 &  &  & \\
2014/10/21 13:35 & 2014/10/21 13:36 & 2014/10/21 13:37 & 2014/10/21 13:37 & M1.2 & S14E35 & 74 & 14.0 & 0.87 & 11.3 & 122.7 & -174.1\\
2014/10/22 01:16 & 2014/10/22 01:36 & 2014/10/22 01:46 & 2014/10/22 01:49 & M8.7 & S13E21 & 733 & 19.7 & 3.41 & 73.2 & 317.5 & -843.2\\
2014/11/05 09:26 & 2014/11/05 09:38 & 2014/11/05 09:42 & 2014/11/05 09:46 & M7.9 & N20E68 & 514 & 18.7 & 3.68 &  &  & \\
2014/11/15 11:40 & 2014/11/15 11:48 & 2014/11/15 11:58 & 2014/11/15 12:02 & M3.2 & S09E63 & 866 & 16.2 & 1.65 &  &  & \\
2014/11/15 20:38 & 2014/11/15 20:40 & 2014/11/15 20:43 & 2014/11/15 20:45 & M3.7 & S13E63 & 322 & 16.0 & 1.93 &  &  & \\
2015/01/13 04:13 & 2015/01/13 04:15 & 2015/01/13 04:21 & 2015/01/13 04:24 & M5.6 & N06W70 & 528 & 16.2 & 2.89 &  &  & \\
2015/03/12 04:41 & 2015/03/12 04:41 & 2015/03/12 04:43 & 2015/03/12 04:45 & M3.2 & S15E11 & 246 & 19.0 & 1.51 & 28.0 & 292.4 & -369.0\\
2015/03/12 21:44 & 2015/03/12 21:46 & 2015/03/12 21:48 & 2015/03/12 21:50 & M2.7 & S15E01 & 252 & 18.0 & 1.34 & 9.7 & 258.2 & -352.4\\
2015/03/15 09:36 &  &  &  & M1.0 & S20W24 &  &  &  &  &  & \\
2015/06/22 17:23 & 2015/06/22 17:49 & 2015/06/22 17:58 & 2015/06/22 18:00 & M6.5 & N12W08 & 639 & 19.1 & 2.38 & 24.7 & 520.0 & -511.8\\
2015/06/25 08:02 & 2015/06/25 08:04 & 2015/06/25 08:14 & 2015/06/25 08:14 & M7.9 & N09W42 & 594 & 18.3 & 2.66 & 13.7 & 242.7 & -190.5\\
2015/09/29 19:08 & 2015/09/29 19:22 & 2015/09/29 19:23 & 2015/09/29 19:24 & M1.1 & S20W36 & 117 & 12.5 & 0.76 & 16.3 & 75.5 & -77.8\\
2015/09/30 13:14 & 2015/09/30 13:18 & 2015/09/30 13:19 & 2015/09/30 13:20 & M1.1 & S23W59 & 62 & 15.8 & 0.74 & 6.2 & 221.2 & -156.9\\
2015/10/01 13:03 & 2015/10/01 13:05 & 2015/10/01 13:09 & 2015/10/01 13:10 & M4.5 & S23W64 & 311 & 17.9 & 2.27 &  &  & \\
2015/10/02 00:06 & 2015/10/02 00:07 & 2015/10/02 00:10 & 2015/10/02 00:13 & M5.5 & S19W67 & 349 & 18.1 & 2.71 &  &  & \\
\enddata
\end{deluxetable}

\floattable
\begin{deluxetable}{ccccccrccccc}
\tablecaption{NWL event list and physical parameters from \textit{GOES} \& \textit{SDO} data. \label{tbl2}}
\rotate
\tablecolumns{12}
\tablenum{2}
\tablewidth{0pt}
\tablehead{
\colhead{} & \colhead{} & \colhead{} & \colhead{} & \colhead{GOES} & \colhead{} & \colhead{derivative} & \multicolumn{2}{c}{@derivative end} & \colhead{$1600{\rm \AA}$ ribbon} & \multicolumn{2}{c}{HMI field strength} \\
\cline{8-9}
\colhead{GOES flare start} & \multicolumn{3}{c}{GOES derivative} & \colhead{X-ray} & \colhead{Sunspot} & \colhead{duration} & \colhead{Temperature} & \colhead{EM} & \colhead{distance} & \multicolumn{2}{c}{@$1600{\rm \AA}$ ribbon} \\
\cline{2-4}\cline{11-12}
\colhead{\scriptsize{YYYY/MM/DD hh:mm}} & \colhead{start} & \colhead{peak} & \colhead{end} & \colhead{class} & \colhead{location} & \colhead{[sec]} & \colhead{[MK]} & \colhead{[$10^{49}/{\rm cm}^3$]} & \colhead{[$\times 10^3 {\rm km}$]} & \colhead{positive [G]} & \colhead{negative [G]}
}
\startdata
2012/03/05 02:30 & 2012/03/05 03:30 & 2012/03/05 03:45 & 2012/03/05 03:55 & X1.1 & N19E58 & 1483 & 18.7 & 4.55 & 33.0 & 539.2 & -523.3\\
2012/07/12 15:37 & 2012/07/12 16:11 & 2012/07/12 16:32 & 2012/07/12 16:44 & X1.4 & S13W03 & 2011 & 17.9 & 6.47 & 46.7 & 975.8 & -436.5\\
2013/05/15 01:24 & 2013/05/15 01:25 & 2013/05/15 01:43 & 2013/05/15 01:48 & X1.2 & N12E64 & 1366 & 16.3 & 6.57 &  &  & \\
2014/10/25 16:55 & 2014/10/25 16:47 & 2014/10/25 17:03 & 2014/10/25 17:07 & X1.0 & S10W22 & 1214 & 19.8 & 4.79 & 40.4 & 521.6 & -383.5\\
2014/11/07 16:53 & 2014/11/07 17:17 & 2014/11/07 17:22 & 2014/11/07 17:25 & X1.6 & N14E36 & 475 & 17.5 & 7.83 & 26.6 & 399.4 & -413.7\\
2011/02/16 07:35 & 2011/02/16 07:37 & 2011/02/16 07:40 & 2011/02/16 07:43 & M1.1 & S19W29 & 358 & 16.1 & 0.58 & 23.3 & 521.3 & -519.7\\
2011/09/23 21:54 & 2011/09/23 21:59 & 2011/09/23 22:03 & 2011/09/23 22:10 & M1.6 & N12E56 & 700 & 13.3 & 0.88 & 58.2 & 274.8 & -890.5\\
2011/11/03 10:58 & 2011/11/03 10:59 & 2011/11/03 11:07 & 2011/11/03 11:10 & M2.5 & N20E70 & 647 & 16.8 & 1.26 &  &  & \\
2011/11/05 03:08 & 2011/11/05 03:29 & 2011/11/05 03:30 & 2011/11/05 03:32 & M3.7 & N20E47 & 197 & 14.9 & 2.00 & 27.3 & 492.4 & -345.4\\
2011/11/05 20:31 & 2011/11/05 20:33 & 2011/11/05 20:35 & 2011/11/05 20:37 & M1.8 & N21E37 & 258 & 17.2 & 0.89 & 11.5 & 391.1 & -327.5\\
2011/12/31 16:16 & 2011/12/31 16:21 & 2011/12/31 16:24 & 2011/12/31 16:25 & M1.5 & S25E42 & 262 & 15.6 & 0.85 & 27.1 & 248.6 & -353.7\\
2012/01/17 04:41 & 2012/01/17 04:43 & 2012/01/17 04:45 & 2012/01/17 04:52 & M1.0 & N18E53 & 547 & 13.8 & 0.59 & 11.3 & 199.2 & -140.5\\
2012/01/18 19:04 & 2012/01/18 19:06 & 2012/01/18 19:09 & 2012/01/18 19:11 & M1.7 & N17E32 & 305 & 14.6 & 0.91 & 11.1 &  & \\
2012/05/07 14:03 & 2012/05/07 14:05 & 2012/05/07 14:19 & 2012/05/07 14:26 & M1.9 & S20W49 & 1224 & 14.1 & 1.00 & 41.1 & 166.8 & -241.4\\
2012/05/09 14:02 & 2012/05/09 14:04 & 2012/05/09 14:06 & 2012/05/09 14:08 & M1.8 & N06E22 & 280 & 14.8 & 1.04 & 20.7 & 223.2 & -594.1\\
2012/06/13 11:29 &  &  &  & M1.2 & S16E18 &  &  &  &  & \\
2012/07/06 08:17 & 2012/07/06 08:18 & 2012/07/06 08:22 & 2012/07/06 08:23 & M1.5 & S17W40 & 319 & 15.4 & 0.84 & 13.7 & 809.5 & -291.3\\
2012/07/14 04:51 & 2012/07/14 04:51 & 2012/07/14 04:54 & 2012/07/14 04:58 & M1.0 & S22W36 & 366 & 13.8 & 0.59 & 44.8 & 575.1 & -623.0\\
2013/05/02 04:58 & 2013/05/02 04:59 & 2013/05/02 05:04 & 2013/05/02 05:08 & M1.1 & N10W26 & 553 & 13.8 & 0.61 & 23.4 & 341.9 & -116.9\\
2013/05/03 16:39 & 2013/05/03 16:40 & 2013/05/03 16:46 & 2013/05/03 16:49 & M1.3 & N10W38 & 529 & 13.1 & 0.72 & 15.0 & 140.5 & -166.1\\
2013/08/17 18:16 & 2013/08/17 18:19 & 2013/08/17 18:22 & 2013/08/17 18:23 & M3.3 & S07W30 & 275 & 16.5 & 1.74 & 7.9 & 386.8 & -260.6\\
2013/10/22 00:14 & 2013/10/22 00:15 & 2013/10/22 00:18 & 2013/10/22 00:21 & M1.0 & N06E17 & 356 & 15.0 & 0.55 & 25.3 & 390.9 & -490.7\\
2013/10/28 14:46 & 2013/10/28 14:55 & 2013/10/28 14:59 & 2013/10/28 15:00 & M2.7 & S08E28 & 286 & 16.6 & 1.22 & 4.5 & 318.6 & -221.2\\
2013/12/07 07:17 & 2013/12/07 07:19 & 2013/12/07 07:22 & 2013/12/07 07:28 & M1.2 & S16W49 & 522 & 12.0 & 0.81 & 34.1 & 117.2 & -171.1\\
2013/12/22 08:05 & 2013/12/22 08:07 & 2013/12/22 08:09 & 2013/12/22 08:11 & M1.9 & S20W49 & 235 & 16.5 & 0.99 & 28.2 & 365.2 & -183.4\\
\enddata
\end{deluxetable}

\floattable
\begin{deluxetable}{ccccccrccccc}
\tablecaption{NWL event list and physical parameters from \textit{GOES} \& \textit{SDO} data. (cont.)}
\rotate
\tablecolumns{12}
\tablenum{2}
\tablewidth{0pt}
\tablehead{
\colhead{} & \colhead{} & \colhead{} & \colhead{} & \colhead{GOES} & \colhead{} & \colhead{derivative} & \multicolumn{2}{c}{@derivative end} & \colhead{$1600{\rm \AA}$ ribbon} & \multicolumn{2}{c}{HMI field strength} \\
\cline{8-9}
\colhead{GOES flare start} & \multicolumn{3}{c}{GOES derivative} & \colhead{X-ray} & \colhead{Sunspot} & \colhead{duration} & \colhead{Temperature} & \colhead{EM} & \colhead{distance} & \multicolumn{2}{c}{@$1600{\rm \AA}$ ribbon} \\
\cline{2-4}\cline{11-12}
\colhead{\scriptsize{YYYY/MM/DD hh:mm}} & \colhead{start} & \colhead{peak} & \colhead{end} & \colhead{class} & \colhead{location} & \colhead{[sec]} & \colhead{[MK]} & \colhead{[$10^{49}/{\rm cm}^3$]} & \colhead{[$\times 10^3 {\rm km}$]} & \colhead{positive [G]} & \colhead{negative [G]}
}
\startdata
2014/01/04 10:16 & 2014/01/04 10:18 & 2014/01/04 10:20 & 2014/01/04 10:24 & M1.3 & S05E48 & 350 & 14.2 & 0.74 & 47.3 & 136.1 & -240.2\\
2014/01/04 19:05 & 2014/01/04 19:04 & 2014/01/04 19:15 & 2014/01/04 19:28 & M4.0 & S11E34 & 1450 & 14.7 & 1.87 & 109.7 &  & \\
2014/02/14 02:40 & 2014/02/14 02:47 & 2014/02/14 02:54 & 2014/02/14 02:56 & M2.3 & S12W25 & 547 & 15.3 & 1.24 & 16.4 & 390.8 & -275.7\\
2014/02/14 12:29 & 2014/02/14 12:30 & 2014/02/14 12:39 & 2014/02/14 12:40 & M1.6 & S15W36 & 580 & 14.0 & 0.91 & 30.9 & 138.5 & -118.4\\
2014/06/12 21:34 & 2014/06/12 21:40 & 2014/06/12 21:57 & 2014/06/12 22:03 & M3.1 & S20W55 & 1413 & 13.7 & 1.37 & 65.9 & 589.9 & -180.8\\
2014/06/15 23:50 & 2014/06/15 23:52 & 2014/06/15 23:56 & 2014/06/15 23:59 & M1.0 & S19E08 & 432 & 12.1 & 0.62 & 36.3 & 121.8 & -187.4\\
2014/08/01 17:55 & 2014/08/01 17:56 & 2014/08/01 18:00 & 2014/08/01 18:05 & M1.5 & S10E11 & 543 & 12.2 & 0.84 & 39.0 & 388.6 & -146.5\\
2014/10/26 19:59 & 2014/10/26 19:58 & 2014/10/26 20:05 & 2014/10/26 20:12 & M2.4 & S15W45 & 848 & 16.7 & 1.06 &107.0 & 315.0 & -768.5\\
2014/10/28 02:15 & 2014/10/28 02:11 & 2014/10/28 02:37 & 2014/10/28 02:39 & M3.4 & S14W61 & 1685 & 16.0 & 1.75 &  &  & \\
2014/10/29 09:54 & 2014/10/29 09:55 & 2014/10/29 09:58 & 2014/10/29 10:01 & M1.2 & S18W77 & 365 & 14.4 & 0.69 &  &  & \\
2014/11/05 18:50 & 2014/11/05 19:24 & 2014/11/05 19:31 & 2014/11/05 19:39 & M2.9 & N17E65 & 900 & 15.0 & 1.48 &  &  & \\
2014/11/07 02:01 & 2014/11/07 02:39 & 2014/11/07 02:42 & 2014/11/07 02:46 & M2.7 & N17E50 & 452 & 16.2 & 1.34 & 28.9 & 564.0 & -427.6\\
2014/11/07 09:43 & 2014/11/07 10:13 & 2014/11/07 10:17 & 2014/11/07 10:21 & M1.0 & N15E43 & 428 & 14.3 & 0.55 & 9.7 & 317.4 & -267.6\\
2014/12/04 07:36 & 2014/12/04 08:01 & 2014/12/04 08:05 & 2014/12/04 08:08 & M1.3 & S24W27 & 403 & 13.9 & 0.69 & 17.7 & 470.3 & -177.8\\
2014/12/05 11:33 &  &  &  & M1.5 & S23W41 &  &  &  &  &  & \\
2014/12/19 09:31 & 2014/12/19 09:33 & 2014/12/19 09:39 & 2014/12/19 09:43 & M1.3 & S19W27 & 619 & 13.1 & 0.79 & 31.3 & 244.1 & -219.3\\
2015/03/12 12:09 & 2015/03/12 12:08 & 2015/03/12 12:11 & 2015/03/12 12:13 & M1.4 & S18E05 & 314 & 12.3 & 0.90 & 68.5 & 481.5 & -380.5\\
2015/03/12 13:45 & 2015/03/12 14:02 & 2015/03/12 14:04 & 2015/03/12 14:08 & M4.2 & S15E06 & 361 & 17.4 & 2.07 & 24.0 & 417.0 & -298.8\\
2015/03/13 05:49 & 2015/03/13 06:00 & 2015/03/13 06:03 & 2015/03/13 06:07 & M1.8 & S14W02 & 385 & 16.0 & 0.96 & 10.4 & 317.5 & -400.0\\
2015/03/14 04:23 & 2015/03/14 04:33 & 2015/03/14 04:36 & 2015/03/14 04:40 & M1.3 & S14W12 & 383 & 14.4 & 0.77 & 39.8 & 144.4 & -655.1\\
2015/03/15 22:42 & 2015/03/15 22:45 & 2015/03/15 22:46 & 2015/03/15 22:47 & M1.2 & S19W32 & 131 & 14.2 & 0.20 & 55.0 & 304.0 & -586.6\\
2015/03/16 10:39 & 2015/03/16 10:41 & 2015/03/16 10:50 & 2015/03/16 10:56 & M1.6 & S17W39 & 876 & 15.1 & 0.86 & 28.9 & 165.5 & -342.2\\
2015/03/17 22:49 & 2015/03/17 23:28 & 2015/03/17 23:30 & 2015/03/17 23:32 & M1.0 & S21W56 & 270 & 12.0 & 0.66 & 43.4 & 107.7 & -106.4\\
2015/06/21 01:02 & 2015/06/21 01:22 & 2015/06/21 01:27 & 2015/06/21 01:37 & M2.0 & N12E13 & 895 & 15.2 & 0.94 & 29.3 & 803.1 & -681.8\\
2015/08/30 02:01 & 2015/08/30 02:54 & 2015/08/30 02:56 & 2015/08/30 03:01 & M1.4 & S17W80 & 445 & 13.9 & 0.61 &  &  & \\
2015/09/27 20:54 & 2015/09/27 20:55 & 2015/09/27 20:57 & 2015/09/27 21:00 & M1.0 & S21W16 & 287 & 14.6 & 0.57 & 10.8 & 378.9 & -286.7\\
2016/01/01 23:10 & 2016/01/01 23:30 & 2016/01/01 23:37 & 2016/01/01 23:44 & M2.3 & S25W82 & 811 & 13.0 & 0.95 &  &  & \\
\enddata
\end{deluxetable}

\begin{table}
\tablenum{3}
\begin{center}
\caption{WLF \& NWL event list and physical parameters from \textit{RHESSI} data. \label{tbl3}}
\begin{tabular}{cccrrrr}
\tableline\tableline
GOES flare start & GOES & Sunspot & Derivative & $50-100~{\rm keV}$ HXR peak & Power law & Energy deposition rate\\
\scriptsize{YYYY/MM/DD hh:mm} &  X-ray class & location & duration [s] & [${\rm counts/s/cm^2/keV}$] & index & ($>30~{\rm keV}$) [erg/s]\\
\tableline
NWL events\\
2013/05/15 01:24 & X1.2 & N12E64 & 1366 & 0.53 & -3.2 & 3.09E+27\\
2014/10/25 16:55 & X1.0 & S10W22 & 1214 & 0.07 & -8.3 & 5.97E+28\\
2011/12/31 16:16 & M1.5 & S25E42 & 262 & 0.12 & -2.7 & 2.56E+26\\
2013/05/02 04:58 & M1.1 & N10W26 & 553 & 0.67 & -2.8 & 1.02E+27\\
2013/05/03 16:39 & M1.3 & N10W38 & 529 & 0.12 & -3.9 & 8.04E+26\\
2013/10/22 00:14 & M1.0 & N06E17 & 356 & 0.06 & -4.5 & 5.46E+26\\
2013/12/22 08:05 & M1.9 & S20W49 & 235 & 0.07 & -2.0 & 3.58E+25\\
2014/01/04 10:16 & M1.3 & S05E48 & 350 & 0.15 & -3.6 & 1.28E+27\\
2014/02/14 12:29 & M1.6 & S15W36 & 580 & 0.04 & -2.4 & 6.20E+25\\
2015/03/15 22:42 & M1.2 & S19W32 & 131 & 0.21 & -6.6 & 4.35E+25\\ \hline
WLF events\\
2011/02/15 01:44 & X2.2 & S20W10 & 631 & 0.74 & -4.3 & 1.34E+28\\
2012/10/23 03:13 & X1.8 & S13E58 & 170 & 7.67 & -3.1 & 3.66E+28\\
2014/10/22 14:02 & X1.6 & S14E13 & 850 & 2.93 & -4.7 & 9.36E+28\\
2014/10/24 20:50 & X3.1 & S16W21 & 783 & 1.03 & -5.9 & 1.06E+29\\
2014/10/27 14:12 & X2.0 & S17W52 & 1237 & 0.26 & -6.2 & 3.85E+28\\
2015/03/11 16:11 & X2.1 & S17E22 & 512 & 64.2 & -7.1 & 3.77E+28\\
2011/02/18 09:55 & M6.6 & S21W55 & 233 & 0.43 & -3.0 & 1.84E+27\\
2011/12/31 13:09 & M2.4 & S25E46 & 254 & 0.11 & -3.4 & 3.97E+26\\
2012/05/09 12:21 & M4.7 & N13E31 & 268 & 0.12 & -3.4 & 7.44E+26\\
2012/05/10 20:20 & M1.7 & N12E12 & 263 & 0.20 & -4.6 & 5.10E+27\\
2012/07/05 03:25 & M4.7 & S18W29 & 90 & 0.34 & -4.7 & 3.16E+27\\
2012/07/06 01:37 & M2.9 & S18W41 & 156 & 1.27 & -3.3 & 4.18E+27\\
2013/10/28 15:07 & M4.4 & S06E28 & 315 & 0.18 & -3.3 & 1.14E+27\\
2013/12/22 14:45 & M3.3 & S19W56 & 278 & 0.08 & -3.1 & 3.47E+26\\
2014/10/22 01:16 & M8.7 & S13E21 & 733 & 2.15 & -3.7 & 2.38E+28\\
2015/03/12 04:41 & M3.2 & S15E11 & 246 & 1.02 & -5.1 & 1.20E+28\\
2015/03/12 21:44 & M2.7 & S15E01 & 252 & 0.47 & -4.5 & 6.92E+27\\
\tableline
\end{tabular}
\end{center}
\end{table}

\end{document}